\documentclass[a4paper,fleqn]{cas-dc}

\usepackage[numbers]{natbib}

\def\tsc#1{\csdef{#1}{\textsc{\lowercase{#1}}\xspace}}
\tsc{WGM}
\tsc{QE}
\tsc{EP}
\tsc{PMS}
\tsc{BEC}
\tsc{DE}
\begin{document}
\let\WriteBookmarks\relax
\def\floatpagepagefraction{1}
\def\textpagefraction{.001}
\shorttitle{BE$\alpha p$R database}
\shortauthors{J.C.~Batchelder and A.M.~Hurst}

\title [mode = title]{The Online Berkeley Evaluated Alpha and Proton Radioactivity database (BE$\alpha p$R) --- from Li to Og!}                      
%\tnotemark[1,2]

%\tnotetext[1]{This document is the results of the research
%   project funded by the National Science Foundation.}

\author[1]{J. C. Batchelder}[type=editor,
%                        auid=000,bioid=1,
%                        prefix=Sir,
%                        role=Researcher,
                        orcid=0000-0003-4629-7586]
%\cormark[1]
\ead{batchelder@berkeley.edu}
\ead[url]{https://nucleardata.berkeley.edu/research/betap.html}

\credit{Evaluation, Data curation, Writing - Original draft preparation}

%\address[1]{, Street 129, 1043 NX Amsterdam, The Netherlands}
\affiliation[1]{organization={Department of Nuclear Engineering, University of California},
                %addressline={}, 
                city={Berkeley},
%               citysep={}, % Uncomment if no comma needed between city and postcode 
                state={California},
                postcode={94720},
                country={USA}}

\author[1]{A. M. Hurst}[orcid=0000-0001-8183-8187]
\credit{Website development, Writing - Original draft preparation}

%\cortext[cor2]{Principal corresponding author}
%\fntext[fn1]{This is the first author footnote, but is common to third
%  author as well.}
%\fntext[fn2]{Another author footnote, this is a very long footnote and
%  it should be a really long footnote. But this footnote is not yet
%  sufficiently long enough to make two lines of footnote text.}

%\nonumnote{This note has no numbers. In this work we demonstrate $a_b$
%  the formation Y\_1 of a new type of polariton on the interface
%  between a cuprous oxide slab and a polystyrene micro-sphere placed
%  on the slab.
%  }

\begin{abstract}
  We have developed the new Berkeley Evaluated Alpha and proton Radioactivity (BE$\alpha p$R) database detailing information on all known $\beta^{+}$-delayed and direct heavy-particle emitters (p, $\alpha$, cluster, fission).  The compiled data encompasses branching ratios, half lives, and all relevant $Q$-values and particle-separation energies.  These are listed for all nuclei where such decays are energetically possible.  In addition, for nuclei with known discrete proton and alpha transitions, the particle-emission energies, intensities, and associated initial and final states are also provided.  A graphical illustration of the major decay modes and a corresponding list of experimental references is also given for each dataset.  The compiled nuclides are organized by their isospin projection ($T_{z}$) and split into even and odd $Z$ datasets. Complete compilations of the database from $T_{z} = -4$ to $T_{z} = +31$ are included.  We intend to maintain this database and post updates periodically as new relevant data becomes available.  In its current construct, information from this database can currently be downloaded, in part or entirely, in portable document format as we are developing a more useful machine-readable format for future dissemination.  Making the database publicly available in the manner described here represents the first step in the ongoing development of a fully open-source project.
\end{abstract}


\maketitle

\section{Introduction to heavy charged-particle emission}

Nuclei far from stability reveal properties of nuclear structure phenomena characterized an extreme imbalance of neutron to proton number with respect to stable nuclei, allowing for a better understanding of fundamental nuclear interactions. In most cases, the study of heavy charged-particle (defined as heavier than a $\beta$ particle) decay modes is the only method available that can populate the nuclear states necessary to provide insight into these interactions.

Nuclei near the proton drip line with large $Q$ values often $\beta^{+}$ decay to excited states that subsequently decay by the emission of a proton (or alpha particle). This is known as beta-delayed proton (or alpha) emission ($\beta^{+}_{p}$ or $\beta^{+}_{\alpha}$). It is a typical decay mode of very neutron-deficient nuclei. Valuable information associated with the ground state in the precursor, such as half life ($T_{1/2}$), spin ($J$), and parity ($\pi$), can be obtained by studying the properties of these decays. The high efficiency and unique experimental signature for detecting heavy charged particles allows for the study of states populated in the daughter nucleus following $\beta^{+}$-decay that are otherwise inaccessible. By measuring the properties of particles emitted to a known state in the daughter, information on the structure of the particle-unbound state can be obtained.

Beta-delayed proton emission is a two step process in which a proton-rich precursor nucleus $\beta^{+}$ decays into a state in an intermediate nucleus, hereafter referred to as the emitter, that is proton unbound. Proton decay from this unbound state occurs rapidly, so the overall $T_{1/2}$ is characteristic of the $\beta^{+}$-decay $T_{1/2}$. This type of decay is energetically possible when the mass of the parent precursor ($A$, $Z$) is larger than the mass of the $\beta^{+}_{p}$ daughter ($A-1$, $Z-2$) plus the mass of the emitted proton:
\begin{equation*}
M_{\text{parent}}(A,Z) > (M_{\text{daughter}}(A-1,Z-2) + m_{p}).
\end{equation*}
 Beta-delayed proton emission has been observed in every element from C ($Z=6$) to Lu ($Z=71$).  Beta-delayed $\alpha$ emission occurs in the same fashion, with many known examples from C ($Z=6$) to Cs ($Z=51$). 

For nuclei that are energetically open to heavy charged-particle emission ($p$, $\alpha$, cluster), direct emission is a competing decay mode.  In direct emission, the charged-particle can tunnel through the Coulomb and centrifugal barriers. Observation of these particles from direct emission not only allows us to establish the limits of stability for a given element, but also provides information on the structure and mass of the parent nucleus. 

\begin{figure*}
	\centering
	\includegraphics[width=.9\textwidth]{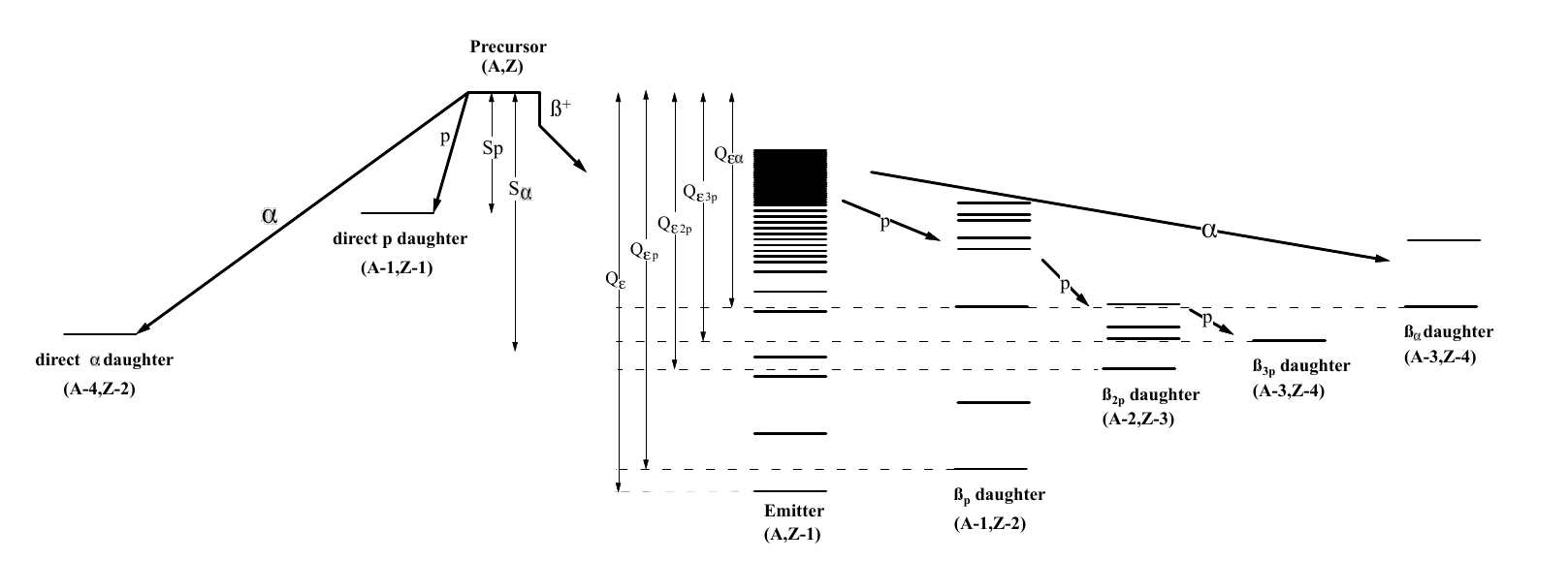}
	\caption{Hypothetical schematic representation of possible decay modes associated with an isotope beyond the proton drip line.}
	\label{FIG:1}
\end{figure*}

These aforementioned decay modes are illustrated in Figure~\ref{FIG:1} depicting a precursor nucleus that is unbound to direct and $\beta^{+}$-delayed particle emission.  In proton-rich lighter nuclei $\beta^{+}$-delayed particle emission tends to dominate, whereas direct proton emission becomes increasingly likely as the $N:Z$ ratio diminishes further.  For high-$Z$ nuclei above $N=84$, $\alpha$ emission becomes the dominant decay mode. 

The Berkeley Evaluated Alpha and proton Radioactivity database (BE$\alpha p$R) is a horizontal evaluation of experimentally-observed decay properties associated with all known $\beta^{+}$-delayed and direct heavy charged-particle emitters ($p$, $\alpha$, cluster and fission) encompassing the entire Segr\'e chart from Li ($Z=3$) to Og ($Z=118$) and is disseminated online \cite{BEApR}.  The information provided in BE$\alpha p$R includes branching ratios, $T_{1/2}$, and all relevant $Q$ values and separation energies $S$ (either taken directly, mostly from Ref. \cite{2021Wa} except where noted, or calculated) listed for those nuclei where these decays are energetically possible. In addition, for those nuclei with measured proton and alpha transitions, the particle energies and intensities, as well as the energies of the corresponding particle-emitting states, are also compiled. A list of experimental references for each precursor is also given in the relevant data table. 

This database grew from a previous publication cataloguing all known $\beta^{+}$-delayed proton and $\alpha$ emitters \cite{2020Bat}. Since then, new measurements have reported more precise values for branching ratios, energies, and other quantities -- information that a static publication cannot be revised to reflect. This motivated the creation of the online BE$\alpha p$R database, which can be kept current and remains useful to its users over time. It incorporates all information from the previous work and extends it to include nuclei that undergo direct heavy charged-particle emission ($p$, $2p$, $\alpha$, cluster, and fission).

The nature of a horizontal evaluation gathers all information associated with the topic in one place, allowing users to explore patterns and trends in the data which can lead to the discovery of new physical phenomena.  This database \cite{BEApR} provides a comprehensive up-to-date compilation of $\beta^{+}$ and $\beta^{+}$-delayed charged-particle emitters that is organized by isospin projection ($T_{z} = (N-Z)/2$).  Organizing the data tables in this way is instructive because nuclides defined by a common $T_{z}$ value tend to have similar properties.  For the cases of $\beta^{+}$-delayed particle emission, nuclei have similar properties across $T_{z}$ chains.  For example, in the case of even-$Z$ $T_{z} = -3/2$ chain, $\beta^{+}$-decay primarily proceeds through the Isobaric Analog State (IAS) followed by proton emission (see $^{57}$Zn \cite{2022Sa}, $^{61}$Ge \cite{2007Bl}, $^{65}$Se \cite{2011Ro}, $^{69}$Br \cite{2011Ro}, and $^{73}$Sr \cite{2019Si}).  Also, $\alpha$-emitting nuclei decay along a unique $T_{z}$ chain and the sequence can be represented within the same dataset.
  
Our database contains complete evaluations from $T_{z} = -4$ to $T_{z} = +31$, arranged according to even-$Z$ and odd-$Z$, giving rise to running total of 1251 nuclei so far.  Isomers characterized by $T_{1/2} > 10$~ns are treated as separately.  Only decays from these “long-lived” states are included, not those from very short-lived high-energy states that emit $p$ or $\alpha$ particles.  No attempt is made for this database to include theoretical predictions or theoretical references.  Information from this database can currently be downloaded in portable document format as we are developing a more useful machine-readable format for future dissemination.  This database is updated periodically as new papers are published, allowing for more rapid dissemination over more traditional nuclear structure databases.

\section{Purpose of database and how it differs from other available databases}

The overall purpose of the BE$\alpha p$R database is to aid researchers on the topic of heavy charged-particle emission.  It can also serve as a pedagogical tool that helps demonstrate, for example, the relationship between energy and branching ratios as well as competition between different decay modes.  The goal of the database is to serve as an authoritative reference for all experimentally known heavy charged-particle radioactivity data: direct and $\beta^{+}$-delayed $\alpha$ decay; direct and $\beta^{+}$-delayed proton emission (including two- and three-proton emission cases); cluster decay; spontaneous and $\beta^{+}$-delayed fission. Note that for fission, only the branching ratio is reported, and not detailed information on the associated daughter nuclides.

Rather than requiring researchers to search the journal literature and databases whose primary purpose differs from the topic, BE$\alpha p$R consolidates all measurements hitherto from the discovery of $\alpha$-decay into a consistent and standardized format.  All values are presented in uniform energy units (MeV) in the center-of-mass reference frame, with well-defined uncertainty notation, and any values that are derived from mass systematics are clearly flagged. Calculated quantities such as hindrance factors in $\alpha$ decay are included to enable direct comparison between theory and experiment across the nuclear chart.  This database is designed to serve both experimentalists and theorists: (i) Experimentalists can use it to benchmark new measurements, identify decays that are poorly known or yet to have been measured, and place new results in context.  (ii) Theorists working on nuclear structure, decay models, or astrophysical reaction rates can extract the systematic trends needed to test their calculations and constrain their models. The combination of exhaustive coverage, primary-source traceability, and up-to-date literature makes it a living reference document rather than a static snapshot.

\begin{figure*}[t]
	\centering
	\includegraphics[width=\textwidth]{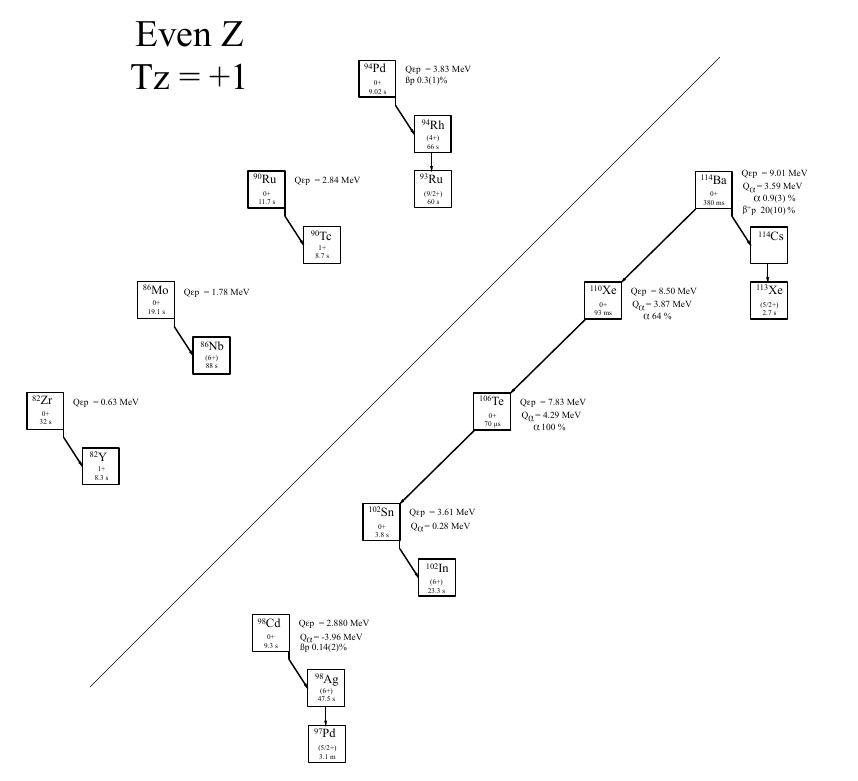}
	\caption{Decay diagram from the even-$Z$, $T_{z} =+1$ dataset.}
	\label{FIG:2}
\end{figure*}

Because datasets are categorized according to $T_{z}$ projections, the BE$\alpha p$R database is organized along $\alpha$-decay chains, thereby allowing the user to selectively retrieve information on the entire decay sequence.  This method of organization is in stark contrast to the Evaluated Nuclear Structure Data File (ENSDF) \cite{ENSDF} which arranges nuclei around $\beta$-decay chains.  Consequently, this requires users to selectively query several different datasets in ENSDF where complete information on the entire $\alpha$-decay chain is required.  In addition, $\beta$-decay products in ENSDF are updated infrequently and usually as part of an overall mass-chain evaluation, which may render the currency of any particular $\beta$-decay chain, or nuclides within a chain, to be out of date with respect to recent experimental developments.  In fact, it is not uncommon for ENSDF to be running several years behind if results of the latest experimental measurements have yet to to be incorporated into an evaluation.

\begin{figure*}[t]
	\centering
	\includegraphics[width=\textwidth]{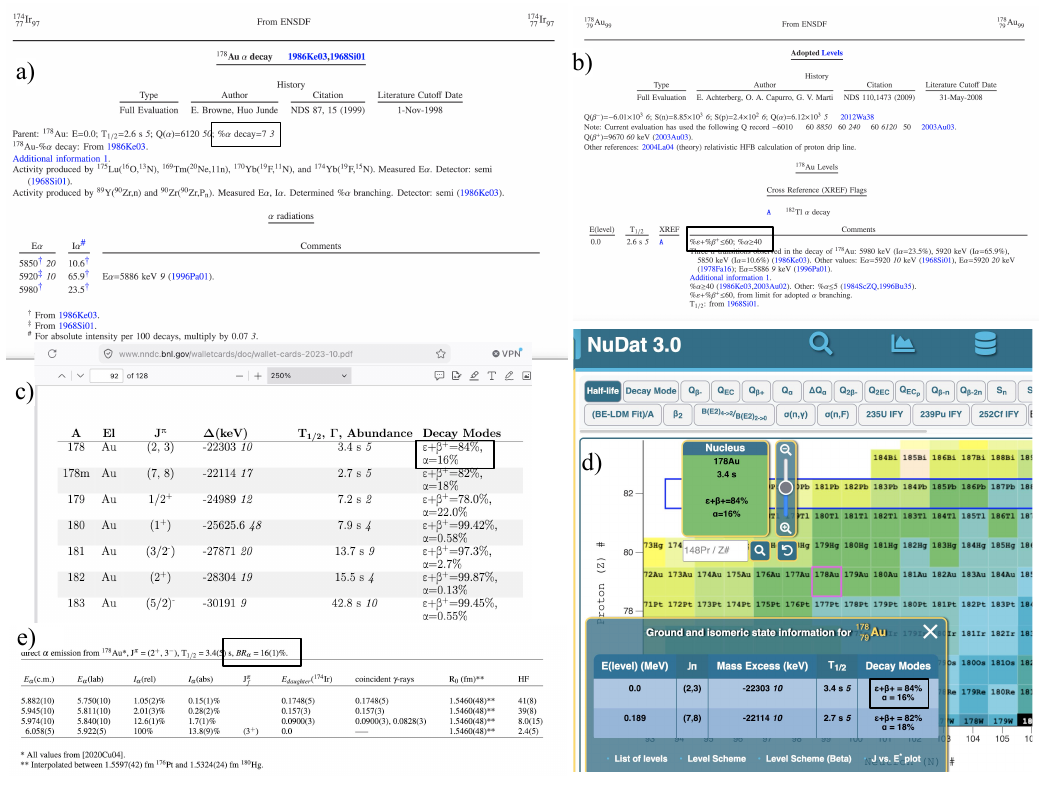}
	\caption{Comparision of results in the various online databases for the ground-state $\alpha$ decay ($I_{\alpha}$) of $^{178}$Au: (a) $I_{\alpha}=7(3)\%$, taken from the ENSDF $\alpha$ decay \cite{1999Br};  (b) $I_{\alpha} \geq 40\%$, taken from the ENDSF adopted levels \cite{2009Ac}; (c) $I_{\alpha} = 16\%$, taken from the Nuclear Wallet Cards \cite{Wallet}; (d) $I_{\alpha} = 16\%$, taken from NuDat \cite{NuDat}; (e) $I_{\alpha} = 16(1)\%$, is the current entry in BE$\alpha p$R.  Boxes are drawn around the relevant information to highlight these differences.  Note that there are three different values listed in the NNDC databases for the $^{178}$Au $\alpha$-decay branching ratio.}
	\label{FIG:3}
\end{figure*}

For example, the decay of $^{22}$Si was last updated in ENSDF in 2015 (both adopted levels and decay databases) \cite{2015Ba, 2015Fi}, detailing only $\beta$-delayed $1p$ emission with a branching ratio of 32(4)\%.  In the meantime, three papers \cite{2022Ci, 2020Le, 2017Xu} have been published offering new information on both $\beta$-delayed $1p$ and $2p$ emission with branching ratios of 61.8(52)\% and 0.7(3)\% respectively. These new results have changed our understanding of this important drip-line nucleus.  A similar example is the decay of $^{178}$Au, last evaluated and updated in ENSDF in 2009 with an $\alpha$-decay branching ratio of $\ge$40\% (adopted levels) \cite{2009Ac}, and in 1999 with an $\alpha$-decay branching ratio of 7(3)\% (decay) \cite{1999Br}.  Since then, new measurements have been published with new $\alpha$-emission transitions and improved accuracy for both the $T_{1/2}$ and $\alpha$-decay transition energies of this nuclide \cite{2026An, 2021Gi, 2020Cu}. In addition, a different $\alpha$-decay branching ratio measurement of 16(1)\% is reported in Ref.~\cite{2020Cu}. Figure~\ref{FIG:3} shows screen captures of the $^{178}$Au information contained in ENSDF (adopted levels and decay) \cite{2009Ac, 1999Br}, the Nuclear Wallet Cards \cite{Wallet} and NuDat \cite{NuDat},\footnote{Screen captures taken June 5, 2026.} illustrating how these databases reflect outdated evaluations, whereas BE$\alpha p$R incorporates all recent measurements and provides the most up-to-date information for this nuclide.

These cases are far from unique -- many similar examples exist throughout ENSDF.  Since evaluations of adopted levels and decay data in ENSDF are often performed by different evaluators at different times, inconsistencies between reported values for the same nucleus may naturally arise.  This is exemplified in Fig.~\ref{FIG:3}(a) and (b).  Table~\ref{tbl1} compares values from BE$\alpha p$R, ENSDF (both adopted and decay datasets) and NuDat for all the Au isotopes contained in the BE$\alpha p$R database, with significant discrepancies in branching ratios and half-lives highlighted in bold.  Such discrepancies are not limited to the gold iostopes, however; similar inconsistencies can be found throughout the Segr{\'e} chart.

%\begin{table*}[width=0.9\linewidth,cols=5,pos=h]
\begin{table*}[h]
\caption{Comparison of branching ratios and half-lives in the BE$\alpha$pR, ENSDF and NuDat databases.}\label{tbl1}
\begin{tabular*}{\tblwidth}{@{} LLLLLL@{} }
\toprule

Nuclide& Quantity & BE$\alpha p$R & ENSDF (adopted) & ENSDF (decay) & NuDat (wallet cards)\\
$^{169}$Au &\%p&$\approx$ 100\% \cite{2019Uu01}&{\bf-----}&{\bf-----}&{\bf-----}\\
&T$_{1/2}$&$<$ 5 $\mu$s \cite{2019Uu01}&{\bf-----}&{\bf-----}&{\bf-----}\\\hline

$^{170}$Au &\%p&89(10)\% \cite{2004Ke06}&89(10)\% \cite{2018Ba41}& 89(10)\% \cite{2026Ba05}&89\%\\
&T$_{1/2}$&286$^{+50}_{-40}$ $\mu$s  \cite{2004Ke06}&290$^{+50}_{-40}$ $\mu$s &290$^{+50}_{-40}$ $\mu$s&290$^{+50}_{-40}$ $\mu$s\\
$^{170m}$Au &\%p&58(5)\% \cite{2004Ke06}&58(5)\%&58(5)\% &58\%\\
&T$_{1/2}$&617$^{+50}_{-40}$ $\mu$s \cite{2004Ke06}&620$^{+50}_{-40}$ $\mu$s&620$^{+50}_{-40}$ $\mu$s&620$^{+50}_{-40}$ $\mu$s\\\hline

$^{171}$Au &\%p&100\% \cite{2004Ke06}&100\%  \cite{2018Ba33}& 100\% \cite{ 2018Ba41}&100\%\\
&T$_{1/2}$&22$^{+3}_{-2}$ $\mu$s  \cite{2004Ke06}&22$^{+3}_{-2}$ $\mu$s &{\bf17$^{+9}_{-5}$ $\mu$s }&22.0$^{+30}_{-20}$ $\mu$s\\
$^{171m}$Au &\%p&40(4)\%  \cite{2004Ke06,1997Da07}&40(6)\%  &46(4)\%&40\%\\
&T$_{1/2}$&1.09(3) ms  \cite{2004Ke06}&1.04(3) ms &1.02(10) ms&1.036(34) ms\\\hline

$^{172}$Au &\%$\alpha$& 100\%  \cite{2009Ha42}&100\%  \cite{ENSDFnow}&100\% \cite{2010Ba27}&100\%\\
&T$_{1/2}$&22$^{+6}_{-4}$ ms  \cite{2009Ha42}&22$^{+6}_{-4}$ ms&22$^{+6}_{-4}$ ms&22$^{+6}_{-4}$ ms\\
$^{172m}$Au &\%$\alpha$&100\%  \cite{2009Ha42}&$\approx$ 100\% &100\% &$\approx$ 100\%\\
&T$_{1/2}$&5(1) ms  \cite{2009Ha42,1996Pa01,1993Se09}&{\bf7.7(14) ms}&{\bf7.7(14) ms}&{\bf8.2$^{+0.9}_{-0.8}$} \\\hline

$^{173}$Au &\%$\alpha$&94$^{+6}_{-19}$ \%  \cite{2012Th13}&94(19)\%  \cite{ENSDFnow}&94\% \cite{2026Ba05}&94\%\\
&T$_{1/2}$&26.3(12) ms  \cite{2012Th13}&26.3(12) ms &25.5(10) s& 26.3(12) ms\\
$^{173m}$Au &\%$\alpha$&92$^{+8}_{-13}$ \%  \cite{2012Th13}&92(13)\% &92\%&92\%\\
&T$_{1/2}$&12.2(1) ms  \cite{2012Th13}&12.2(1) ms &12.2(1) ms&12.2(1) ms \\\hline

$^{174}$Au &\%$\alpha$&90(6)\%  \cite{2002Ro17}&{\bf$>$ 0}  \cite{1999Br24} &90(6)\%\cite{2018Ba41}&90\%\\
&T$_{1/2}$&139(3) ms  \cite{2004GoZZ}&{\bf120(20) ms} &139(3) ms&139(3) ms\\
$^{174m}$Au &\%$\alpha$&obs  \cite{2004GoZZ}&-----&$<$ 100\%  \cite{2018Ba41}&$>$0\%\\
&T$_{1/2}$&162(3) ms  \cite{2004GoZZ}&-----&162.9(16) ms&162(3) ms\\\hline

$^{175}$Au &\%$\alpha$&90(7)\%  \cite{2013An10}&90(7)\%  \cite{2025Wu19} &90(7)\%  \cite{2018Ba33}&90\%\\
&T$_{1/2}$&200(3) ms  \cite{2017Ba46}&201(3) ms &201(3) ms&201(3) ms\\
$^{175m}$Au &\%$\alpha$&  90(3)\%  \cite{2010An01}&90(3)\%&90\%&90\%\\
&T$_{1/2}$&137(1) ms  \cite{2017Ba46, 2011Wa37}&137(1)ms&137(1) ms&137(1) ms\\\hline

$^{176}$Au &\%$\alpha$&58(5)\%  \cite{2021Ha32}&{\bf?}  \cite{2006Ba16}&{\bf75(8)\%}  \cite{ENSDFnow}&58\%\\
&T$_{1/2}$&1.046(11) s  \cite{2004GoZZ}&1.05(1) s&1.05(1) s &1.05(1) s\\
$^{176m}$Au &\%$\alpha$&29(5)\%  \cite{2021Ha32}&{\bf-----}&100\% &29\%\\
&T$_{1/2}$&1.36(2) s  \cite{2004GoZZ}&1.36(2) s&1.36(2) s &1.36(2) s\\\hline

$^{177}$Au&\%$\alpha$&64(5)\%  \cite{2021Ha32}&{\bf40(6)\%}  \cite{2019Ko26}&{\bf40(6)\%}  \cite{ENSDFnow}&{\bf54\%}\\
&T$_{1/2}$& 1.501(20) s  \cite{2025Sp01}&1.501(20) s&1.53(7) s&1.454(46) s\\
$^{177m}$Au &\%$\alpha$&56(8)\%  \cite{2021Ha32}&{\bf60(10)\%}&{\bf66(10)\%} &{\bf58\%}\\
&T$_{1/2}$&1.186(12) s  \cite{2021Ha32, 2001Ko14}&1.193(13) s&{\bf1.00(20) s}&1.180(12) s\\\hline

$^{178}$Au&\%$\alpha$&16(1)\%  \cite{2020Cu04}&{\bf$\ge$40\%} \cite{2009Ac01}&{\bf7(3)\%}  \cite{1999Br24}&16\%\\
&T$_{1/2}$&3.4(5) s  \cite{2020Cu04}&{\bf2.6(5) s}&{\bf2.6(5) s}&3.4(5) s\\
$^{178m}$Au &\%$\alpha$&18(1)\%  \cite{2020Cu04}&{\bf-----}&{\bf-----}&{\bf82\%}\\
&T$_{1/2}$&2.7(5) s  \cite{2020Cu04}&{\bf-----}&{\bf-----}&2.7(5)\%\\\hline

$^{179}$Au&\%$\alpha$&22.0(9)\%  \cite{1986Ke03}&22.0(9)\%  \cite{2009Ba02}&{\bf88.0(9)\%}  \cite{2025Wu19}&22.0\%\\
&T$_{1/2}$&7.3(3) s  \cite{2021Ha32}&7.1(3) s&7.3(3) s &7.2(2) s\\\hline

$^{180}$Au& \%$\alpha$&0.58(10)\%  \cite{2020Ha24}&{\bf$\ge$ 1.8\%}  \cite{2015Mc03} &{\bf$>$ 1.8\%}  \cite{2006Ba16}&0.58\%\\
&T$_{1/2}$&8.1(3) s  \cite{1977Hu05}&{\bf8.4(6) s}&8.1(3) s&{\bf7.9(4) s}\\\hline

$^{181}$Au &\%$\alpha$&2.7(5)\%  \cite{1995Bi01}&2.7(5)\%  \cite{2005Wu06}&2.7(5)\%  \cite{2019Ko26}&2.7\%\\
&T$_{1/2}$&14.5(4) s  \cite{1995Bi01}&{\bf13.7(14) s}&{\bf13.7(14) s}&{\bf13.7(9) s}\\\hline

$^{182}$Au&\%$\alpha$&0.129(11)\%  \cite{2025Mi06}&0.13(5)\%  \cite{2015Si18}& 0.13(5)\%  \cite{2009Ac01}&0.13\%\\
&T$_{1/2}$&16.43(12) s  \cite{2025Mi06}&{\bf15.5(4) s}&{\bf15.5(4) s}&{\bf15.5(4) s}\\\hline

$^{183}$Au &\%$\alpha$&0.8(2)\% \cite{1995Bi01} &{\bf0.55(25)\%}  \cite{1983Ba19}&{\bf0.55(25)\%}  \cite{2009Ba02}&{\bf0.55\%}\\
&T$_{1/2}$ &44.6(19) s  \cite{1995Bi01}&{\bf42.8(10) s} &{\bf42.8(10) s}&{\bf42.8(10) s}\\\hline
$^{184m}$Au&\%$\alpha$&0.013(3)\%  \cite{1995Bi01}& {\bf$\leq$0.016\%}  \cite{2010Ba05}& {\bf$\leq$0.016\%}  \cite{2015Mc03} &{\bf$<$0.016\%}\\
&T$_{1/2}$&46.4(10) s  \cite{1995Bi01, 1977Za03, 1992Ro21}& 47.6(14) s&47.6(14) s&46.4(10) s\\\hline

$^{185}$Au &\%$\alpha$&0.26(6)\%  \cite{1995Bi01}&0.26(6)\%  \cite{2005Wu07}&0.26(6)\%  \cite{2005Wu06}&0.26\%\\
&T$_{1/2}$&4.2(1) m  \cite{1995Bi01}&4.25(6) m& 4.25(6) m &4.3(1) m\\\hline
$^{186}$Au&\%$\alpha$&8(2)$\times$10$^{-4}$ \%  \cite{1990Ak04}&8(2)$\times$10$^{-4}$ \%  \cite{2022Ba26}& 8(2)$\times$10$^{-4}$ \%  \cite{2015Si18}&8$\times$10$^{-4}$ \%\\
&T$_{1/2}$& 10.7(5) m  \cite{1970Jo02}&10.7(5) m&10.7(4) m&10.7(4) m\\ \hline
$^{187}$Au &\%$\alpha$&$\approx$ 2$\times$10$^{-3}$ \%  \cite{1968Si01}&{\bf0.003\% sys} \cite{2009Ba12}&{\bf-----}&{\bf?}\\
&T$_{1/2}$& 8.3(3)s  \cite{1983Ga01, 1979Be51}&8.3(2) m&-----&8.3(2) m \\
\bottomrule
\end{tabular*}
\end{table*}

Furthermore, at the time of this writing, many of the more exotic decays contained in BE$\alpha p$R are missing in ENSDF. The interactive web application NuDat \cite{NuDat} sources the Nuclear Wallet Cards \cite{Wallet} for the half lives and branching ratios for a given decay.  However, no explicit references are provided, nor do the Nuclear Wallet Cards give detailed information on individual transitions.  Table~\ref{tbl2} compares the number of nuclei listed by decay mode according to $\beta_{2p}$, $\beta_{3p}$, direct $2p$, and cluster emission (known clusters include C, O, F, Ne, Mg, and Si) in BE$\alpha p$R versus those given in ENSDF (adopted levels and decay) and NuDat.  Note that $\beta$-delayed particle emission listed here is a combination of $\beta^{+}$ and electron capture, i.e., $\beta^{+}/\epsilon$, as very few of the measurements detailed in BE$\alpha p$R are able to distinguish between these two modes.

%\begin{table*}[width=0.9\linewidth,cols=5,pos=h]
%\begin{table*}[h]
\begin{table}[h]
\caption{Comparison of the BE$\alpha$pR, ENSDF and NuDat databases for the more exotic decay modes that have been observed in proton-rich nuclei.  The cluster-decay modes encompass C, O, F, Ne, Mg, and Si particle emission.}\label{tbl2}
\begin{tabular*}{\tblwidth}{@{} LLLLL@{} }
\toprule
Decay&BE$\alpha$pR & ENSDF (adopted) & ENSDF (decay) & NuDat\\
\midrule
$\beta_{2p}$   &14 & 10 & 8 & 11 \\
$\beta_{3p}$&4 & 2 & 3 & 4 \\
$2p$&13&13&5&13\\
cluster&17&13&1&17\\
\bottomrule
\end{tabular*}
\end{table}
%\end{table*}

Our database gives the user an up-to-date resource for all available information on proton-rich heavy charged-particle emitters.  We would also like to emphasize that all energies, branching ratios, and half-life values contained in BE$\alpha p$R are taken from the original sources, not from other evaluations, and are extensively referenced.

Another invaluable resource that we have leveraged in the creation of the BE$\alpha p$R database is \textit{"The AME 2020 atomic mass evaluation"} (AME2020) \cite{2021Wa}.  Most of the $Q$-values and particle-separation energies that we include are taken from, or derived from, AME2020 \cite{2021Wa}.  However, AME2020 alone does not allow us to capture all information reported with respect to these quantities.  Notable exceptions to this are instances where we have derived information from $\alpha$- and proton-decay energies; such occurrences are clearly noted and referenced.  Although the AME2020 is a phenomenal piece of work on masses of nuclei across the entirety of the nuclear landscape, it does not contain complete information on individual charged-particle decays to excited states that we also include in BE$\alpha p$R.  

Finally, the  BE$\alpha p$R database contains a complete list of targeted references for each nucleus.  To be considered for inclusion here, the reference must contain relevant information regarding the charged-particle decay of the nucleus in question.  This is far more selective and very different from the Nuclear Science References (NSR) \cite{NSR} which lists all papers that simply mention the nucleus (even if the paper contains no information related to the decay of the nucleus in question) and also lists papers that identify the specified nucleus as a possible parent or daughter in radioactive decay.  Accordingly, a simple query on the nucleus of interest in NSR will likely retrieve many papers not directly relevant to the topic area, thus, requiring the reader to filter through individual results to find the appropriate set of reference documentation.

\section{Database contents and the website}

The database contains all measured and predicted $Q$-values, separation energies, half lives, branching ratios, and individual transition energies based on information from Ref. \cite{2021Wa} other than cases where a more accurate value can be obtained from the particle energies (typically from newer papers).  It is currently available online \cite{BEApR}.  From the landing page, users can browse by individual isospin projection ($T_{z}$) or by element ($Z$) and mass ($A$) number.  In addition, there is an option to download the entire database, as well as summarized tables of direct-particle emitters and $\beta^{+}$-delayed particle emitters, in portable document format.

For each individual $T_{z}$ dataset, the first page provides a decay schematic showing the relevant $Q$ values and separation energies, decay modes, and $T_{1/2}$ values for each nucleus in the $T_{z}$ chain, along with the date of the most recent update.  An example illustrating the decay chain for $T_{z} = +1$, even-$Z$, is shown in Figure~\ref{FIG:2}.  Accompanying each figure is a series of tables that are organized as follows:
\begin{itemize}
\item Table 1 contains the $T_{1/2}$ and $J^{\pi}$ information for each nucleus along with all the $\beta^{+}$-delayed information:$Q_{\beta_{x}}$ values (where defines the charged-particle emission, e.g., $x=p,2p,3p,\alpha$), experimental branching ratios and associated references.   
\item Table 2 contains all of the direct particle-emission information: $S_{p}$, $S_{2p}$, $Q_{\alpha}$ values, experimental branching ratios and associated references.
\item Table 3 (and beyond) contains individual transitions associated with $\beta^{+}$-delayed or direct particle-emission decay modes --- where known --- for a particular nucleus belonging to the $T_{z}$ chain.  Higher-rank tables, needed for any additional nuclides in the chain containing individual transitions, are arranged in order from lightest-$A$ to heaviest-$A$ in the respective $T_{z}$ chain.
\end{itemize}
In Tables 1 and 2 of of the BE$\alpha p$R database documentation, the primary reference(s) is presented in bold.  Evaluator comments are provided with the tables as necessary.  The information given in these datasets allows users to explore relationships between various reported quantities to gain a better handle on the underlying physics involved.

In its current construct, the BE$\alpha$pR database \cite{BEApR} contains a total of 139 $T_{z}$ decay chains, comprising 1463 $\beta^{+}$-delayed and direct charged-particle emission decay modes from 1251 nuclei.  Additionally, and unique to this database, we have collated 3986 discrete charged-particle transitions into BE$\alpha p$R.  As it stands, there are 4287 unique references associated with the compiled information.  

\section{Summary}

As experimental programs at facilities worldwide continue to push the boundaries of nuclear stability, BE$\alpha p$R \cite{BEApR} is positioned to serve as the definitive living reference for heavy charged-particle radioactivity data for both experimentalists and theorists in nuclear structure and nuclear astrophysics.  The BE$\alpha p$R database provides a comprehensive overview of all experimentally-known heavy charged-particle decay data from both direct and $\beta^{+}$-delayed charged-particle $p$, $\alpha$, cluster, and fission decay modes.  The database provides associated decay information for all nuclei known to exhibit these decay modes, spanning the whole of the nuclear landscape from Li to Og, organized by isospin projection from $T_{z} = -4$ to $+31$. The arrangement of $T_{z}$-chain organization provides complete $\alpha$-decay sequence information in a single dataset that cannot otherwise be obtained in a convenient and expedient manner.  The BE$\alpha p$R database also contains entries for decay modes underrepresented in other databases, such as ENSDF \cite{ENSDF}, including $\beta^{+}_{2p}$, $\beta^{+}_{3p}$, direct $2p$ and cluster emission, while other databases, such as AME2020 \cite{2021Wa}, do not contain comprehensive fine-structure information on these charged-particle decays to individual excited states.

The database is openly disseminated online \cite{BEApR}, where users can interact and browse by $T_{z}$ chain, by $Z$ and $A$, or download the complete database into a single file.  In the future, we plan to update information periodically to ensure that BE$\alpha p$R remains topical as new experimental results are published.  We consider this to be a critical feature of the proposed maintenance work given the rapid pace of discovery in the physics of exotic nuclei at and beyond the proton drip line.  Finally, inline with previous modernization efforts carried out by the group \cite{2023Hu, PyPI:pyEGAF, 2023Hu:b, PyPI:paceENSDF}, we are currently developing a new machine-readable format for this database together with an associated API that will allow for convenient interaction, manipulation and bespoke visualization of the data.

\section*{Acknowledgments}

This work was supported by the Lawrence Berkeley National Laboratory under Contract No. DE-AC02-05CH11231, and by the University of California, Berkeley, through the U.S. Department of Energy National Nuclear Security Administration through the Nuclear Science and Security Consortium under Award Number DE-NA0003996.

%\appendix
%\section{My Appendix}
%Appendix sections are coded under \verb+\appendix+.

%\verb+\printcredits+ command is used after appendix sections to list 
%author credit taxonomy contribution roles tagged using \verb+\credit+ 
%in frontmatter.

\printcredits

%\bibliographystyle{cas-model2-names}
\bibliographystyle{elsarticle-num}

\bibliography{BEApR_refs}


\end{document}